\documentclass[onecolumn,showpacs,prb]{revtex4}
\usepackage{graphicx}
\usepackage{amsmath}
\usepackage{latexsym}
\def\bphi{\mbox{\boldmath$\phi$}}
\begin{document}
\title{ Soliton solutions of a gauged O(3) sigma model
with interpolating potential}
\author{Pradip Mukherjee}
\altaffiliation{mukhpradip@gmail.com}
\altaffiliation{Also Visiting Associate, S. N. Bose National Centre for Basic Sciences, JD Block, Sector III, Salt Lake City, Calcutta -700 098, India}
\affiliation{Department of Physics, Presidency College,\\86/1 College Street, Kolkata-700073,  India.}

\begin{abstract}
Soliton modes in a gauged sigma model
with interpolating potential have been investigated. Numerical solutions using a fourth order Runge -- Kutta method are discussed. By tuning the interpolation parameter the transition from symmetrybreaking to the symmetric phase is highlighted.
\end{abstract}
\pacs{11.10.Kk; 11.10.Lm; 11.15.-q}
\maketitle
\section{Introduction}
    The  O(3) nonlinear sigma model has long been the subject of intense research due to its theoretical and phenomenological basis. This theory describes classical (anti) ferromagnetic spin systems at their critical points in Euclidean space, while in
the Minkowski one it delineates the long wavelength limit of quantum antiferromagnets. The model exhibits solitons,
Hopf instantons and novel spin and statistics in 2+1 space-time dimensions with inclusion of the Chern-Simons term.

 The
soliton solutions of the model exhibit scale invariance which poses difficulty in the particle interpretation on quantization. A popular means of breaking this scale invariance is to gauge a U(1) subgroup of the O(3) symmetry of the model by coupling the
sigma model fields with a gauge field through the corresponding U(1)
current. {\footnote{This is different from the minimal coupling via the topological
current discussed previously \cite{wil}.}} This class of gauged O(3) sigma models in three dimensions have
been studied over a long time\cite{{sch},{gho},{lee},{muk1},{muk2},{mend}, {land}}. Initially
the gauge field dynamics was assumed to be dictated by the Maxwell term
\cite{sch}. Later the extension of the model with
the Chern - Simons coupling was investigated \cite{gho}. A particular form of
self - interaction was required to be included in these models in order to 
saturate
the Bogomol'nyi bounds \cite{bog}. The form of the assumed self - interaction
potential is of crucial importance. The minima of the
potential determine the vaccum structure of the theory. The solutions change remarkably
when the vaccum structure exhibits spontaneous breaking of the symmetry
of the gauge group. Thus it was demonstrated that the observed degeneracy
of the solutions of \cite{sch,gho} is lifted when potentials with
symmetry breaking minima were incorporated \cite{muk1,muk2}. The studies of the gauged O(3) sigma model is important due to their intrinsic interest and also due to the fact that the soliton solutions of the gauged
O(3) Chern-Simons model may be relevant in planar condensed matter systems \cite{pani,pani1,han}. Recently gauged 
nonlinear sigma model was considered in order to obtain self-dual cosmic string solutions \cite{verb, ham}. This explains the continuing interest in such models in the literature \cite{{sch},{gho},{lee},{muk1},{muk2},{mend}, {land}}.

  A particular aspect of the gauged O(3) sigma models where the gauge field dynamics is governed by the Maxwell term can be identified by comparing the results of \cite{sch} and \cite{muk2}. In \cite{sch} the vaccum is symmetric and the $n = 1$ soliton solution does not exist,
$n$ being the topological charge. Here, solutions exists for $n = 2$ onwards. Moreover, these soliton solutions have arbitrary magnetic flux. When we achieve symmetrybreaking vaccum by chosing the potential appropriately \cite{muk2} soliton solutions are obtained for $n = 1$. {\footnote{The disappearence of the $n = 1$ soliton has been shown to follow from general analytical method in \cite{sch1}.}}These solutions have quntized magnetic flux and qualify as magnetic vortices. It will be interesting to follow the solutions from the symmetrybreaking to the symmetric phase. This is the motivation of the present paper.

  We will consider a generalisation of the models of \cite{sch} and \cite{muk2}
 with an adjustable real parameter $v$ in the
expression of the self - interaction potential which interpolates between the symmetric and the symmetrybreaking vaccua. This will in particular
 allow us to investigate the soliton solutions in the entire regime of the symmetrybreaking vacuum structures and also to follow the collapse of the $n = 1$ soliton as we move from the assymmetric 
to the symmetric phase. The solitons of the model are obtained as the solutions of the self -- dual equation obtained by saturating the Bogomoln'yi bounds. Unfortunately, these equations fall outside the   Liouville class even after assuming a rotationally symmetric ansatz. Thus
exact analytical solutions are not obtainable and numerical methods are to be invoked.

          The organisation of the paper is as follows. In the following section we present a brief review of the  O(3) nonlinear sigma model. This will be helpful in presenting our work in the proper context. In section 3 our model is introduced. General topological classifications of the soliton solutions of the model has been discussed here. In section 4 the saturation of the self-dual limits has been examined and the Bogomol'nyi equations have been written down. Also the analytical form of the  Bogomol'nyi equations has been worked out assuming a rotationally symmetyric ansatz. These equuations, even in the rotationally symmetric scenario, are not exactly integrable. A numerical solution has been performed to understand the details of the solution. A fourth order Runge -- Kutta algorithm is adopted with provision of tuning the potential appropriately.
In section 5 the numerical method and some results are presented. We conclude in section 6.

\section {  O(3) nonlinear sigma models } 

It will be useful to start with a brief review of the 
  nonlinear O(3) sigma model \cite{bel}. The lagrangian of the model
is given by,
\begin {equation}
{\cal L} =\frac {1} {2}\partial_\mu {\bphi}\cdot 
\partial^\mu
{\bphi}\label{LO3}
\end {equation}
Here ${\bphi}$ is a triplet of scalar fields
constituting a vector in the internal space with unit norm
\begin {eqnarray}
\phi_a = {\bf n_a}\cdot {\bphi}, ~(a=1,2,3)\\
{\bphi}\cdot{\bphi} =\phi_a\phi_a= 1
\label{CONST}
\end {eqnarray}
The vectors ${\bf  n}_a$
constitute a basis of unit orthogonal vectors in the internal space.
We work in the Minkowskian space - time with the metric tensor
diagonal, $g_{\mu\nu} = (1,-1,-1)$.
 
The finite energy solutions of the model (\ref{LO3})
satisfies the boundary condition 
\begin{equation}
 {\rm lim} \phi^a\hspace{.2cm}=\hspace{.2cm}\phi^a_{(0)}\label{BN}
\end{equation}
at physiacal infinity.
The condition (\ref{BN}) corresponds to one point compactification of the
physical infinity. 
The physical space $R_2$ becomes topologically equivalent to $S_2$
due to this compactification. The static finite energy solutions of the
model are then maps  from this sphere to the internal sphere. Such 
solutions are 
classified by the homotopy \cite{dho}
\begin{equation}
\Pi_2(S_2) = Z\label{HOMO}
\end{equation}
We can construct a current
\begin{equation}
K_\mu = {\frac{1}{8\pi}}\epsilon_{\mu\nu\lambda}\bf{\phi}\cdot(
       \partial^\nu\bf{\phi}\times\partial^\lambda\bf{\phi})\label{TCUR}
\end{equation}
which is conserved irrespective of the equation of motion. The corresponding
charge 
\begin{eqnarray}
T &=& \int d^2{\bf x} K_0\nonumber\\
  &=& {\frac{1}{8\pi}}\int d^2{\bf x} \epsilon_{ij}
      \bf{\phi}\cdot (\partial^i\bf{\phi}\times\partial^j\bf{\phi})\label{TCHAR}
\end{eqnarray}
gives the winding number of the mapping (\ref{HOMO}) \cite{raj}.

\section{Our model - topological classification of the soliton solutions}

In the class of gauged models of our interest here 
a U(1) subgroup of the rotation symmetry
of the model (\ref{LO3}) is gauged. We chose this to be
 the SO(2) [U(1)] subgroup of rotations  
about the 3 - axis in the internal space. The Lagrangian of our model is given
by

\begin {equation}
{\cal L} ={\frac{1}{2}}D_\mu {\bphi}\cdot D^\mu
{\bphi}
 -{\frac{1}{4}}F_{\mu\nu}F^{\mu\nu}+ U({\bphi})\label{LGO3}
\end {equation}

$D_\mu {\bphi}$ is the covariant derivative given by
\begin {equation}
D_\mu {\bphi}=\partial_\mu {\bphi}
+ A_\mu {\bf n}_3\times {\bphi}
\end {equation}
 The SO(2)
(U(1)) subgroup is gauged by the vector potential $A_\mu$ whose dynamics is
 dictated
by the Maxwell term. Here $F_{\mu\nu}$ are the electromagnetic field tensor,
\begin {equation}
F_{\mu\nu} = \partial_\mu A_\nu - \partial_\nu A_\mu
\end {equation}
$ U(\bphi)$ is the self - interaction  potential
required for saturating the self - dual limits. We chose
\begin {equation}
U({\bphi})=-{\frac{1}{2}}(v -  \phi_3)^2\label{POT}
\end {equation}
where $v$ is a real parameter. Substituting $v = 0$ we get back the model
of \cite{muk2} whereas $v = 1$ gives the model of \cite{sch}.

    We observe that 
the minima
of the potential arise when ,
\begin {equation}
\phi_3 = v\label{N3B}
\end {equation}
which is equivalent to the condition
\begin {equation}
\phi_1^2+\phi_2^2 =1-v^2\label{NB}
\end {equation}
on account of the constraint (\ref{CONST}).
The values of v must be restricted to
\begin{equation}
{\left |v\right |}\le 1\label{VB}
\end{equation}
The condition (\ref{NB}) denotes a latitudinal circle 
(i.e. circle with fixed latitude) on the unit sphere
in the internal space. 
By varying v from -1 to +1 we span the sphere from the south pole to 
the north pole. It is clear that the finite energy solutions of the model
must satisfy (\ref{NB}) at physical infinity. For $v\ne 1$ this boundary
condition corresponds to the spontaneous breaking of the symmetry of the
gauge group and in the limit
\begin {equation}
{\left |v\right |} \to 1\label{VLIM}
\end{equation}
the asymmetric phase changes to the symmetric phase.
We call the potential (\ref{POT}) interpolating in this sense. 
In the asymmetric phase the soliton solutions are classified according
to the homotopy
\begin{equation}
\Pi_1(S_1) = Z \label{HOM1}
\end {equation}
instead of (\ref{HOMO}). In the symmetric phase, however, this new topology
disappears and the solitons are classified according to (\ref{HOMO})
as in the usual sigma model (\ref{LO3}). 
A remarkable fallout of this change of topology is the disappearence of the soliton with unit charge. The fundamental solitonic
mode $n = 1$ ($n$ being the vorticity) ceases to exist in the symmetric phase.
The modes corresponding to n = 2 onwards still persist but the magnetic flux
associated with them ceases to remain quantised.

  In the asymmetric phase the vorticity is the 
winding number i.e. the number of times by which 
the infinite physical circle winds over the latitudinal circle (\ref{NB}).
Associated with this is a uniqe mapping of the internal sphere where the
degree of mapping is usually fractional. By inspection we 
construct a current
\begin {equation}
K_\mu = {\frac{1} {8\pi}}\epsilon_{\mu\nu\lambda}[
{\bphi}\cdot D^\nu {\bphi}
\times D^\lambda {\bphi} - F^{\nu\lambda}(v - \phi_3)]\label{TCUR1}
\end {equation}
generalising the topological current (\ref{TCUR}). The current (\ref{TCUR1})
is manifestly gauge invariant and differs from (\ref{TCUR}) by the 
curl of a vector field. The conservation principle 
\begin {equation}
\partial_\mu K^\mu = 0
\end {equation}
thus automatically follows from the conservation of (\ref{TCUR}).
The corresponding conserved charge is
\begin {equation}
T = \int d^2x K_0\label{TCH}
\end {equation}
Using (\ref{TCUR1}) and (\ref{TCH}) we can write

\begin {eqnarray}
T &=&\int d^2x[{\frac{1}{8\pi}}\epsilon_{ij}{\bphi}
\cdot(\partial^i{\bphi}
\times \partial^j {\bphi})]\nonumber\\
&-&{ \frac{1}{4\pi}}\int_{boundary}(v -\phi_3) A_\theta r d\theta\label{tnw}
\end {eqnarray}
where r,$\theta$ are polar coordinates in the physical space and $A_\theta
= {\bf e}_\theta \cdot {\bf A}$.
Using the boundary condition (\ref{N3B}) 
we find that T is equal to the degree of the
mapping
of the internal sphere. Note that this situation is different from \cite{lee}
where the topological charge usually differs from the degree of the mapping.
In this context
it is interesting to observe that
 the current (\ref{TCUR1}) is not unique because 
we can always add an arbitrary multiple of
\begin{equation}
{\frac{1}{8\pi}}\epsilon_{\mu\nu\lambda}F^{\nu\lambda}\nonumber
\end{equation}
with it without affecting its conservation. We chose (\ref{TCUR1})
because it generates proper topological charge. 

\section{Self dual equations in the rotationally symmetric ansatz} 

In the previous section we have discussed the general topological classification of the solutions of the equations of motion following from
(\ref{LGO3}). In the present section we will discuss the solution of the equations of motion.
The Euler - Lagrange equations of the system (\ref{LGO3}) is derived subject to the
constraint (\ref{CONST}) by the Lagrange multiplier technique
\begin {eqnarray}
D_\nu (D^\nu {\bphi})& =& [D_\nu (D^\nu
{\bphi})
\cdot {\bphi}] {\bphi} +{\bf n}_3(v - \phi_3)\nonumber\\
&+& (v - \phi_3)\phi_3{\bphi}\label{elphi}\\
\partial_\nu F^{\nu\mu} =j^\mu\label{ela}
\end {eqnarray}
where
\begin {equation}
j^\mu = -{\bf n}_3\cdot{\bf J}^\mu\hspace{.2cm} and\hspace{.2cm}
 {\bf J}^\mu ={\bphi}\times D^\mu {\bphi}\label{jmu}
 \end {equation}
Using (\ref{elphi}) we get
\begin {equation}
D_\mu {\bf J}^\mu = -(v - \phi_3)
({\bf n}_3 \times {\bphi})\phi_3
\end {equation}

 From (\ref{ela}) we find,for static configurations
\begin {equation}
\nabla^2 A^0 = -A^0(1 - \phi_3^2)
\end {equation}
 From the last equation it is evident that we can chose
\begin {equation}
A^0 = 0
\end {equation}
As a consequence we find that the excitations of the model are electrically
neutral.

The equations (\ref{elphi}) and (\ref{ela}) are second order differential equations in time. As is well known, first order equations which are the solutions of the equations of motion can be derived by minimizing the energy functional in the static limit. Keeping this goal in mind
we now construct the
 energy functional  from the symmetric energy - momentum tensor following from (\ref{LGO3}).
The energy
\begin{equation}
E = {\frac{1}{2}}\int d^2{\bf x}\left [ D_0\bphi\cdot D_0\bphi
    - D_i\bphi\cdot D^i\bphi + (v - \phi_3)^2
    - 2(F_0^\sigma F_{0\sigma} - {\frac{1}{4}}F_{\rho\sigma}F^{\rho\sigma})
   \right ].\label{E}
\end{equation}
For static configuration and the choice $A^0$ = 0, $E$ becomes
\begin {equation}
E = {\frac{1}{2}}\int d^2x[(D_i{\bphi})
\cdot(D_i{\bphi})+ F_{12}^2+(v -\phi_3)^2]\label{EST}
\end {equation}
Several observations about the finite energy solutions can be made at
this stage from (\ref{EST}). By defining
\begin{equation}
\psi = \phi_1 + i\phi_2
\end{equation}
we get 
\begin {equation}
D_i {\bphi}\cdot D_i{\bphi}
= |(\partial_i + iA_i)\psi|^2 +(\partial_i\phi_3)^2
\end {equation}
The boundary condition (\ref{NB}) dictates that
\begin {equation}
\psi \approx (1 - v^2)^{{\frac{1}{2}}}e^{in\theta}
\end {equation}
at infinity.
From (\ref{EST}) we observe that for finite energy configurations we
require
\begin {equation}
{\bf{A}}={\bf{e_\theta}}{\frac{n}{r}}\label{AB}
\end {equation}
on the boundary. This scenario is exactly identical with the
observations of \cite{muk2} and leads to the quantisation of 
the magnetic flux
\begin {equation}
\Phi = \int B d^2x = \int_{boundary}A_\theta r d\theta = 2\pi n
\end {equation}
The basic mechanism leading to this quantisation remains operative
so far as $v$ is less than 1. At $v = 1$, however, the gauge field ${\bf A}$
becomes arbitrary on the boundary except for the requirement that
the magnetic field B should vanish on the boundary. Remember
that not all the vortices present in the broaken phase survives this demand.
Specifically, the $n = 1$ vortex becomes inadmissible.  

 Now the search for the self - dual conditions proceed in the usual way.
We rearrange the energy functional as
\begin {equation}
E = {\frac{1}{2}} \int d^{2}x[{\frac{1}{2}}(D_i{\bphi} \pm
\epsilon_{ij}{\bphi}\times  D_{j}{\bphi})^2
+ (F^{12} \mp (v -\phi_3))^2] \pm 4\pi T\label{EBOG}
  \end {equation}
Equation (\ref{EBOG}) gives the Bogomol'nyi conditions
\begin {eqnarray}
D_i{\bphi}\pm  \epsilon_{ij} {\bf\phi}
\times D_j{\bphi} = 0\label{SDPHI}\\
 F_{12}\mp (v -\phi_3) =0\label{SDA} 
 \end {eqnarray}
  which minimize the energy functional in a particular topological sector,
  the upper sign corresponds to +ve and the lower sign corresponds to -ve
  value of the topological charge.

  We will now 
turn towards the analysis of the self - dual equations using the 
rotationally symmetric ansatz \cite{wu}
  \begin {eqnarray}
  \phi_1(r,\theta) = \sin g(r) \cos n\theta\nonumber\\
  \phi_2(r,\theta) = \sin g(r) \sin n\theta\nonumber\\
  \phi_3(r,\theta) = \cos g(r)\nonumber\\
  {\bf A}(r,\theta)= -{\bf e}_\theta {\frac{na(r)}{r}}\label{ANS}
  \end {eqnarray}
  From (\ref{N3B}) we observe that we require the boundary condition
  \begin {equation}
  g(r) \to \cos^{-1}v \hspace{.2cm}
{\rm as}\hspace{.2cm} r \to \infty\label{GB}
  \end {equation}
and equation (\ref{AB}) dictates that
  \begin {equation}
a(r) \to -1 \hspace{.2cm}{\rm as}\hspace{.2cm} r \to \infty\label{aB}
  \end {equation}
Remember that equation (\ref{AB}) was obtained so as the solutions have finite
energy.
  Again for the fields to be well defined at the origin we require
  \begin {equation}
  g(r) \to 0 \hspace{.2cm}{\rm or}\hspace{.2cm}
 \pi \hspace{.2cm}{\rm and} \hspace{.2cm}
  a(r) \to 0\hspace{.2cm} {\rm as} \hspace{.2cm}r \to 0\label{ag0}
  \end {equation}
Substituting the Ansatz(\ref{ANS}) into (\ref{SDPHI}) and (\ref{SDA})
 we find that
  \begin {eqnarray}
  g^\prime (r) = \pm {\frac{n(a+1)}{r}} \sin g,\label{eqg}\\
  a^\prime (r) = \pm {\frac{r}{n}}(v- \cos g)\label{eqa}
  \end {eqnarray}
  where the upper sign holds for +ve T and the lower sign corresponds to
  -ve T.Equations (\ref{eqg}) and (\ref{eqa}) are not exactly integrable.
  In the following section we will discuss the numerical solution 
 of the boundary value problem defined by (\ref{eqg})
and (\ref{eqa}) with (\ref{GB}) to (\ref{ag0}). 

Using the Ansatz (\ref{ANS}) we can explicitly compute the topological charge T
by performing the integration in (\ref{TCH}).The result is
\begin {equation}
T = -{\frac{n}{2}}[\cos g(\infty)-\cos g(0)]-{\frac{1}{2}}[v - \cos g(\infty)]\label{t}
\end {equation}
The second term of (\ref{t}) vanishes due to the boundary condition (\ref{GB}).
Also, when g(0) = 0,
\begin{equation}
T = {\frac{n}{2}}(1 - v)\label{A}
\end{equation}
and, when g(0) = $\pi$,
\begin{equation}
T = -{\frac{n}{2}}(1 + v)\label{B}
\end{equation}
It is evident that T is in general fractional. Due to (\ref{tnw}) it is 
equal to the degree of mapping of the internal sphere. This can also be checked 
explicitly.

 From the above analysis we find that g(0)
= 0
 corresponds to +ve T
and g(0) = $\pi$ corresponds to -ve T.
We shall restrict our attention on negetive T
which will be useful for comparision of results with those available in the literature. The boundary value problem of interest is then
  \begin {eqnarray}
  g^\prime (r) = - {\frac{n(a+1)}{r}} \sin g\label{eqg1}\\
  a^\prime (r) = - {\frac{r}{n}}(v- \cos g)\label{eqa1}
  \end {eqnarray}
with
\begin{eqnarray}
g(0) = \pi,a(0) = 0\nonumber\\
g(\infty) = \cos^{-1}v, a(\infty) = 0\label{BOUN}
\end{eqnarray}
In addition we require $a^\prime (r)$ $\to$ 0 as $r\to \infty$. This condition follows from
(\ref{eqg1}), (\ref{eqa1}) and (\ref{BOUN}) and should be considered as a consistency condition to be satisfied by their soloutions.

\section{Numerical solution}
 The simultaneous equations (\ref{eqg1}) and (\ref{eqa1}) subject to the boundary conditions (\ref{BOUN}) are not amenable to exact solution.
They can however be integrated numerically. We have already mentioned the quenching of the $n = 1$ solution in the limit $v\to 1$.
This is connected with the transition from the symmetry breaking to the symmetric phase. The numerical solution is thus interesting because it will enable us to see how the solutions change as we follow them from the deep assymetric phase $( v = 0)$ to the symmetric phase $(v = 1)$. In the following we provide the results of numerical solution to highlight these issues. 

      Let us note some details of the numerical method. A fourth order Runge -- Kutta method was employed. The point $r=0$ is a regular singular point of the equation. So it was not possible to start the code from $r=0$. Instead, we start it from a small value of $r$. The behaviour of the functions near r = 0 can be easily derived from (\ref{eqg1}) and(\ref{eqa1})
\begin{equation}
g(r)\approx \pi + Ar^n
\label{bc1}
\end{equation}
\begin{equation}
a(r) \approx -{\frac{r^2}{2n}}(1+v)
\label{bc2}
\end{equation}
Here $A$ is an arbitrary constant which fixes the values of g and a at
infinity. In the symmetrybreaking phase the numerical solution depends sensitively on the value of $A$. {\footnote{This should be contrasted with the symmetric vacuum solution where $A$ may have arbitrary values.}}There is a critical value of $A$, $A = A^{crit}$ for which the boundary
conditions are satisfied.
If the value of $A$ is larger than $A^{crit}$
 the conditions at infinity are overshooted,
whereas, if the value is smaller than the critical value 
 g(r) vanishes asymptotically
after reaching a maximum. The situation is comparable with similar
findings elsewhere \cite{jac}.
The values of $g$ and $a$ were calulated at a small value of $r$ using (\ref{bc1}) and (\ref{bc2}). The parameter $A$ was tuned to match boundary conditions at the other end. Interestingly, this matching is not obtainable when $n=1$ and $v =1$. This is consistent with the quenching of the $n = 1$ mode in the symmetric vacuum situation.
       
    After the brief discussion of the numerical technique we will present a summary of the results. As may be recalled, the purpose of the paper is to study the solutions throughout the asymmetric phase with an eye to the disappearence of the $n=1$ mode. Accordingly
profiles of $g$ and $a$ will be given for $n = 1$, for different values of $v$. In figures 1 and 2 these profiles are shown for $v = 0,.2,.4,.6,.8$. The corresponding magnetic field distributions are  given in figure 3. Another interesting issue is the change of the matter and the gauge profiles with the topological charge. In figures 4 and 5 this is demonstrated for  different  $n$ values for a constant  $v$.
\begin{figure}[ht!]
\begin{center}
\includegraphics[width=5.5cm,angle=270]{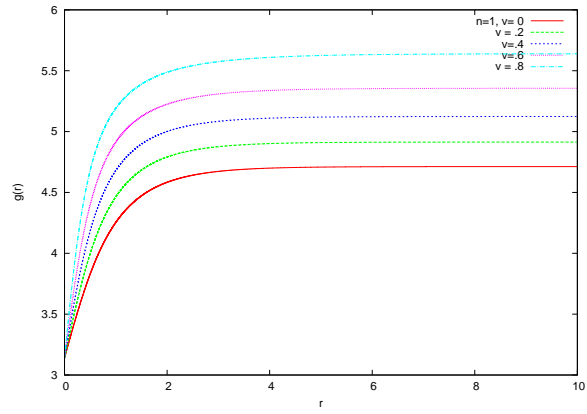}
\caption{\label{fg01} {\it The matter profile.}  The function $g(r)$ is plotted against $r$ for $n = 1$ and different $v$ values, indicated on the right hand top corner.
}
\end{center}
\end{figure}

\begin{figure}[ht1!]
\begin{center}
\includegraphics[width=5.5cm,angle=270]{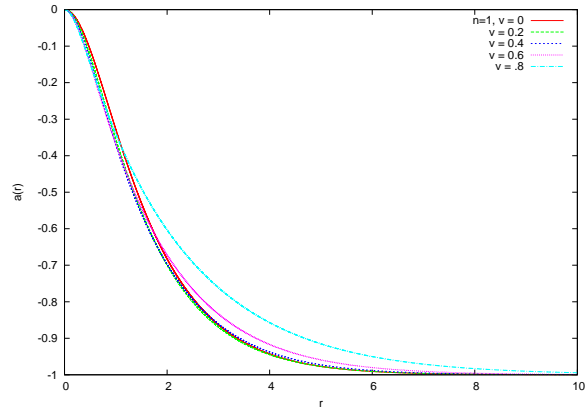}
\caption{\label{fg02} {\it The gauge field profile.}  The function $a(r)$ is plotted against $r$ for $n = 1$ and different $v$ values. 
}
\end{center}
\end{figure}

\begin{figure}[ht2!]
\begin{center}
\includegraphics[width=5.5cm,angle=270]{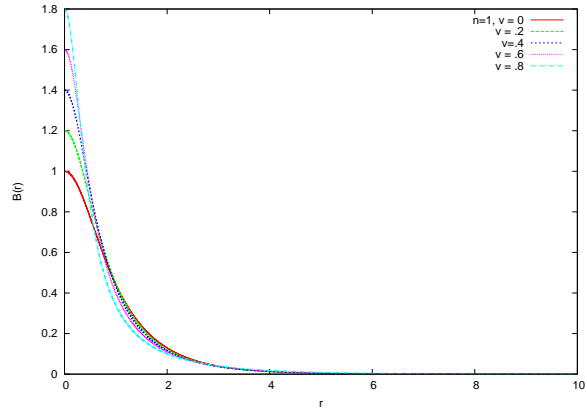}
\caption{\label{fg03} {\it The magnetic field profiles.}  Plot of the function $B(r)$  against $r$.
}
\end{center}
\end{figure}

\begin{figure}[ht4!]
\begin{center}
\includegraphics[width=5.5cm,angle=270]{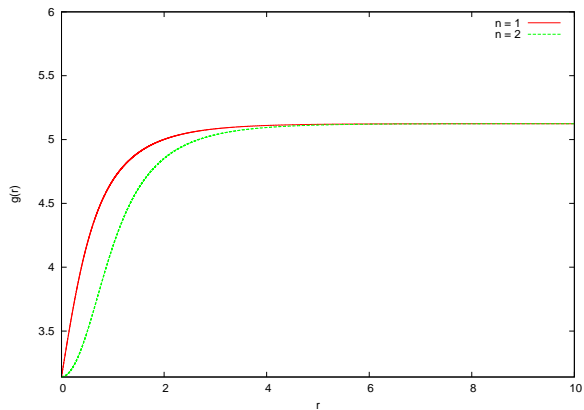}
\caption{\label{fg00} {\it The matter profiles for different n (v = .4).}  
}
\end{center}
\end{figure}

\begin{figure}[ht5!]
\begin{center}
\includegraphics[width=5.5cm,angle=270]{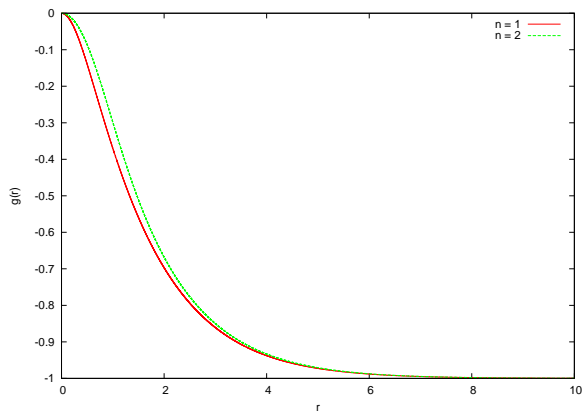}
\caption{\label{fg00} {\it The gauge field profiles for different n (v =.4).} 
}
\end{center}
\end{figure}


\newpage

\section{Conclusion} The O(3) sigma model in (2+1) dimensional space -- time with its U(1) subgroup gauged was mooted \cite{sch} as a possible mechanism to break the scale -- invariance of the soutions of the original 3 - dimensional O(3) sigma model. The model finds possible applications in such diverse areas such as planar condensed matter physics \cite{pani,pani1,han}, gravitating cosmic strings \cite{verb, han} and as such is being continuously explored in the literature \cite{sch,gho,sch1,lee,muk1,muk2,mend,land}. An interesting aspect of the gauged O(3) sigma models is the qualitative change of the soliton modes in the symmetric and symmetrybreaking vacuum scenario, as can be appreciated by a comparision of solutions given in \cite{sch, muk2}. In this paper
 we have considered a gauged O(3) sigma model with the gauge field
dynamics determined by the Maxwell term as in \cite{sch, muk2}. An interpolating potential was included to invesigate the solutions in the entire symmetrybreaking regime
. This potential depends on a free parameter, the
variation of which effects transition from the asymmetric to symmetric phase. We have discussed the transition of the associated topology of the
soliton solutions. The Bogomol'nyi bound was saturated to give the self -- dual solutions of the equation of motion. 
The self - dual equations
are, however, not exactly solvable. They were studied
numerically to trace out the solutions in the entire asymmetric phase with particular emphasis on the $n = 1$ mode. 
Our analysis may be interesting from the point of view of
applications, particularly in condensed matter physics.
\section{Acknowledgement}
The author likes to thank Muktish Acharyya for his assistance in the mumerical solution.


\begin{thebibliography}{99}
\bibitem{sch} B.J. Schroers, Phys. Lett. {\bf{B 356}} (1995) 291.
\bibitem{gho} P.K. Ghosh and S.K. Ghosh, Phys. Lett. {\bf{B366}} (1996) 199.
\bibitem{lee} K. Kimm, K. Lee and T. Lee, Phys. Rev. {\bf{D 53}} (1996) 4436.
\bibitem{muk1} P. Mukherjee, Phys. Lett. {\bf{B 403}} (1997) 70.
\bibitem{muk2} P. Mukherjee, Phys. Rev. {\bf{D 58}} (1998) 105025.
\bibitem{mend}K. C. Mendes, R. R. Landim, C. A. S. Almeida, Mod.Phys.Lett. {\bf{A20}} (2005) 1005
\bibitem{land}  M. S. Cunha, R. R. Landim, C. A. S. Almeida
Phys.Rev. {\bf{D74}} (2006) 067701. 
\bibitem{bog} E.B. Bogomol'nyi, Sov. J. Nucl. Phys. {\bf{24}} (1976) 449.

\bibitem{pani} P. K. Panigrahi , S Roy , and W. Scherer , Phys. Rev. Lett. {\bf 61} (1998) 2827

\bibitem{pani1} P. K. Panigrahi , S. Roy and W. Scherer, Phys. Rev. {\bf D38} (1998) 3199

\bibitem{han} C Han, Phys. Rev. {\bf D47} (1993) 5521.
\bibitem{verb} 
Y. Verbin, S. Madsen, A.L. Larsen,  Phys.Rev. {\bf{D67}} (2003) 085019.
\bibitem{ham}
H. R. Vanaie, N. Riazi, Int.J.Mod.Phys. {\bf{A19}} (2004) 3595. 
\bibitem{bel}  A.A. Belavin and A.M. Polyakov, JETP Lett. {\bf{22}} (1975) 245.

\bibitem{dho} F. J. Hilton, An introduction to Homotopy theory ( Cambridge
University Press, Cambridge, England, 1953 ).
\bibitem{raj} R.Rajaraman, Solitons and Instantons
  ( North Holland Publishing Company, Amsterdam, 1989 ).
\bibitem{sch1} 
B.J. Schroers, Nucl.Phys. {\bf{B475}} (1996) 440.
\bibitem{wu} Y.S. Wu and A. Zee, Phys. Lett. {\bf{B147}} (1984) 325 .
\bibitem{jac} R. Jackiw, K. Lee and E. Weinberg, Phys. Rev.
{\bf{D 42}} (1990) 3488 .
\bibitem{wil}  F. Wilczek and A. Zee, Phys. Rev. Lett. {\bf{51}} (1983) 2250;
M. Bowick, D. Karabali and L. C. R. Wijewardhana, Nucl. Phys. {\bf{B 271}} (1986) 417;
R. Banerjee, Nucl. Phys. {\bf{B 419}} (1994) 611.
\end{thebibliography}
\end{document}